\documentclass[a4paperal]{i3m}

\usepackage{graphicx}
\usepackage{siunitx}
\usepackage{amsmath}
\usepackage{amsfonts}

\conference{emss}

\title{Implementing the BBE Agent-Based Model of a Sports-Betting Exchange}

\author{Dave Cliff, James Hawkins, James Keen, and Roberto Lau-Soto}

\affil{Department of Computer Science, University of Bristol, Bristol BS8 1UB, U.K.}

\authnote{\authfn{1}Authors listed in alphabetic order; please direct correspondence to Dave Cliff, csdtc@bristol.ac.uk}

\runningauthor{D.\ Cliff, J.\ Hawkins, J.\ Keen, \& R.\ Lau-Soto.}

\confyear{2021}

\begin{document}

\begin{frontmatter}
\maketitle
\begin{abstract}
In this paper we describe three independent implementations of a new agent-based model (ABM) that simulates a contemporary sports-betting exchange, such as those offered commercially by companies including {\em Betfair}, {\em Smarkets}, and {\em Betdaq}. The motivation for constructing this ABM, which is known as the {\em Bristol Betting Exchange}\/ (BBE), is so that it can serve as a synthetic data generator, producing large volumes of data that can be used to develop and test new betting strategies via advanced data analytics and machine learning techniques. Betting exchanges act as online platforms on which bettors can find willing counterparties to a bet, and they do this in a way that is directly comparable to the manner in which electronic financial exchanges, such as major stock markets, act as platforms that allow traders to find willing counterparties to buy from or sell to: the platform aggregates and anonymises orders from multiple participants, showing a summary of the market that is updated in real-time. In the first instance, BBE is aimed primarily at producing synthetic data for {\em in-play}\/ betting (also known as {\em in-race}\/ or {\em in-game}\/ betting) where bettors can place bets on the outcome of a track-race event, such as a horse race, after the race has started and for as long as the  race is underway, with betting only ceasing when the race ends. The rationale for, and design of, BBE has been described in detail in a previous paper that we summarise here, before discussing our comparative results which contrast a single-threaded implementation in Python, a multi-threaded implementation in Python, and an implementation where Python header-code calls simulations of the track-racing events written in OpenCL that execute on a 640-core GPU -- this runs  $\approx 1000$ times faster than the single-threaded Python. Our source-code for BBE is being made freely available on GitHub. 
\end{abstract}

\end{frontmatter}

\section{Introduction}
\label{sec:introduction}

Since the rise of web-enabled e-commerce in the late 1990s, the betting industry has been transformed worldwide by the rise of online betting exchanges. Drawing inspiration from the electronic market technologies that had been developed to serve the needs of major financial exchanges, betting exchanges do not act as traditional bookmakers, taking the opposite side of a bet to a customer; instead, a betting exchange operates as an intermediary aggregator platform that enables bettors to efficiently discover counterparties (i.e., other bettors willing to stake some money on the opposing view) and the betting exchange will typically charge a small commission fee to bet-winners for enabling this matching service. This is similar to how financial exchanges enable buyers and sellers of some asset to find counterparties and agree on a fair price, and the similarity between financial exchanges and betting exchanges is sufficiently strong that bettors and exchange operators routinely refer to the distribution of bets over the space of possible outcomes for an event of interest to bettors (such as a horse-race or a soccer or tennis match) as the {\em market}\/ for that event.

The rise of online betting exchanges has enabled styles of betting that were not previously practicable or accessible for most bettors. One example is so-called {\em in-play}\/ betting in which, after an event has started, bettors can continue to make bets about the outcome of that event right up until the end of the event. Another example is often referred to as {\em trading}\/ on the betting exchange, where bettors risk their money not on the outcome of the event itself but instead by gambling on the movement of odds and changes in the distribution of stake-money within the betting exchange's market for that event, right up until it ends. 

Annual revenues for the global gambling industry are currently estimated to be around US\$500billion (\cite{PRNews_2020}) and betting exchanges account for a sizeable proportion of this. Many betting exchanges publish details of application programming interfaces (APIs) which allow third-party developers to write automated systems that interact with the exchange's central systems without needing a human bettor to be interacting with the exchange via a web browser. As has happened in  financial-market exchanges in recent years, the availability of API access to betting exchanges gives rise to the possibility of using artificial intelligence (AI) and machine learning (ML) to create profitable automated betting systems.\footnote{We four authors are all based in the U.K., where gambling as described in this paper is entirely legal, and where gambling-industry companies pay corporate taxes that feed into the budgets for publicly-funded R\&D; we recognise that in other jurisdictions and other cultures gambling is considered morally questionable and/or is illegal, and that this might prompt local questions about the morality of our work as reported here.}

However, many contemporary AI/ML approaches -- most notably Deep Learning Neural Networks (DLNNs: see e.g. \cite{goodfellow_etal_2017})  -- are notoriously data-hungry. That is, they can deliver remarkably successful results, but often to do that they require very large amounts of historical data to be ``trained'' on, to learn from. Very often, acquiring data in sufficient quantities is either very expensive, because the operators of betting exchanges sell their historic data sets at premium rates, or simply impossible (e.g.: you might develop a DLNN system that is obviously going to be profitable, but which requires several hundred year's worth of racing data to be adequately trained -- here the problem is not that you cannot afford the data, but rather that even if you had infinite funds you cannot obtain enough data because the data you require is just not available at any price).

This problem, of AI/ML requiring original data at impracticable scales, is now increasingly addressed by the creation of {\em synthetic data generators} (SDGs; see, e.g., \cite{elemam_etal_2021}). For the purposes of this paper, we'll define an SDG as a generative model that can create new data-sets which preserve the original data's key statistical features, and for which the ground-truths are known and explainable -- that is, for which we know and control the causal mechanistic interactions that led to the generation of the data. The ground-truth requirement means that we are not reliant on inference to determine what set of world conditions gave rise to the data: instead we have a record of what caused the data to be as they are. 

In this paper, we describe our work thus far in implementing a SDG for a contemporary sports exchange, concentrating on the generation of plausible synthetic data-sets for in-play betting on track-racing events, such as horse-races. This paper reports on work-in-progress, and it presents the first comparative empirical results from three independent implementations of an agent-based model of in-play betting on a sports exchange: the agent-based model is named the {\em Bristol Betting Exchange}\/ (BBE). The rationale for developing BBE, the overall architecture of its design, and an extensive literature review, were all recently presented in a long paper by \cite{cliff_2021_bbe}, which this present paper can be considered as a continuation of. In this paper we summarise key points from \cite{cliff_2021_bbe} and then we describe our three independent implementations of BBE, each of which takes a different technical approach, and then we provide comparative results showing BBE in action. Full details of each of the independent implementations are given in \cite{hawkins_2021_MEng,keen_2021_MEng}; and \cite{lausoto_2021_MEng}, and the source-code for the implementations has been made freely available on GitHub as a service to those researchers in our community who wish to replicate or extend our work.\footnote{
GitHub repositories for the three implementations are each available from {\tt https://github.com} in the following paths: \\ 
MCM: {\tt /Yepadee/Bristol-Betting-Exchange}\\
MTP: {\tt /keenjam/BettingExchange} \\
STP: {\tt /RobertoLauSoto/BristolBettingExchange} }  
This paper concentrates on implementation issues in bringing the entire BBE design up to operational capability: this involves creating agent-based models of race events -- the competitors and the bettors, and the betting-exchange matching engine, all of which are described here. Future work will fine-tune the implementations such that the statistical properties of the data synthesized by BBE are in the best possible alignment with those of data from one or more existing commercially-operated betting exchanges.  

Section~\ref{sec:background} provides a brief introduction to betting exchanges and related literature. Then in Section~\ref{sec:racesim} we explain the BBE race simulator; in Section~\ref{sec:exchange} we describe the operation of the BBE betting exchange itself; and in Section~\ref{sec:bettors} we discuss the issue of modelling the bettors who interact with the exchange. All of the content in sections~\ref{sec:background} to~\ref{sec:bettors} is heavily abridged and condensed from \cite{cliff_2021_bbe}: readers familiar with that paper can safely skip straight to Section~\ref{sec:implement} in which we describe the three implementations, and present comparative results. Our plans for future work are discussed in Section~\ref{sec:furtherwork}.

\section{Background: Betting on Exchanges}
\label{sec:background}

Since the dot-com boom of the late 1990's, the worldwide betting industry has transformed from being focused almost entirely on traditional bookmaking as previously practiced for many hundreds of years, to one in which the dominant practice revolves around bettors placing their bets via betting exchanges, and particularly on the internationally successful UK-based betting exchange {\em Betfair}, which is widely credited with being the disruptive innovator in this space, and which rapidly grew to huge financial success. The key innovation in Betfair was recognising that the existence of a population of bettors with varying and opposing views, such that the various possible outcomes of a sporting event each attract some number of bettors willing to either {\em back} some outcome ${\cal O}$ (i.e., to bet that ${\cal O}$ {\em will}\/ happen) or to {\em lay}\/ the outcome (i.e., to bet that $\cal O$ will {\em not}\/ happen, because they believe some other outcome will occur), is very similar to the situation in a financial market where there are some number of traders interested in buying units of an asset, and some number of traders wanting to sell units of the asset. The reason why the services offered by Betfair and similar platforms are described as betting {\em exchanges}\/ is because they bring backers and layers together to identify counterparties to a bet, and give all participants a shared summary view of the distribution of bets for a particular event, in a manner very similar to how most major financial exchanges bring together potential buyers and sellers, give them a summary view of the overall market supply and demand for some tradeable asset, and allow traders in the market to identify counterparties willing to transact at a price agreeable to both parties.

At the heart of a betting exchange for a particular event is a data structure which is referred to as the {\em market}\/ for that event, which is a direct analogue of the {\em limit order book}\/ in a financial exchange (see e.g.\ \cite{gould_etal_2013_LOBs}). For a betting event, the market will typically be displayed to a bettor as graphical user interface (GUI) consisting of a rectangular table or {\em grid}\/ of cells: with each competitor in the race allocated one row on the grid. A betting exchange's market for an event is split between backs and lays, arranged in order of goodness-of-odds, so that each cell in the grid displays a specific odds along with the total amount wagered at those odds. However, the grid display only shows a small number (typically three, sometimes one) of the best prices available to back and lay a particular competitor: for illustration, see \cite{cliff_2021_bbe}. 

Crucially, a betting exchange is not acting like a traditional bookie: it is not carrying risk of losing its own money by laying a customer's back, or backing a customer's lay: instead, it is acting merely as a centralized meeting and matching service for bettors to seek and identify willing counter-parties with whom to bet.

BBE has been specifically designed to model in-play betting. One key aspect of in-play betting that BBE has been designed to explore is the {\em opinion dynamics}\/ within the population of bettors; i.e., the extent to which the opinions of some bettors in the market for an event have their opinions (and hence their subsequent bets) affected by the distribution of money on the market for that event, and by any sudden changes in that distribution, because the distribution of money in the market for an event gives insight into the collective opinions, the overall sentiment, of the population of bettors active in that market. Relevant literature on opinion dynamics is discussed further in \cite{cliff_2021_bbe}.

With a betting exchange publishing details of its in-play market for the various possible outcomes of a particular event comes the opportunity for bettors to risk their money on derivative bets, i.e.\ to not bet directly on the actual outcome of the event, but instead to wager a stake on price-movements within the market while the event is taking place -- this is sometimes referred to as {\em trading}, to distinguish it from betting on the event itself, and is described in more detail below. Traders on betting exchanges often find the limited summary data in the grid-view of the market to be too restrictive, and opt instead for an interface that displays the {\em ladder}\/ for each competitor, a linear display of every available odds/price, and the {\em liquidity} (total sum of staked amounts) at each of those prices.

In this paper, as on most major betting exchanges, all odds will be expressed as decimals (potential total returned), rather than using other representations. For example, where a successful bet with a \$1 stake generates winnings of \$10 plus the original stake returned, for a total of \$11, the decimal representation of the odds is 11; similarly, where a successful bet with a \$5 stake returns winnings of \$1 plus the original stake returned for a total of \$6, the decimal odds are 1.2.

For further discussion of the growth and impact of betting exchanges such as Betfair, see \cite{cliff_2021_bbe}, and the review by \cite{smith_vaughanwilliams_2008}.

Devising profitable automated betting strategies is a labor-intensive activity requiring significant expertise in the design/development phase, and potentially needing access to very large amounts of betting-exchange data, i.e.\ time-series of various betting markets on which strategies can be tested. Betting exchanges do sell such data, but they typically charge premium fees which can be prohibitive for non-professional betting-strategy developers, thereby erecting a major barrier to entry. A primary motivation for the design and development of BBE was to create a source of reliable synthetic data that could be used to explore the application and refinement of AI and ML methods to in-play betting-exchange scenarios, thereby facilitating replacement of the skilled human betting-strategy designer with automated analysis, search and optimization processes: this is returned to in Section~\ref{sec:furtherwork} later in this paper.  

The BBE model as laid out by \cite{cliff_2021_bbe} allows for some sophistication, but in the first instance we are exploring its behavior in minimal, pared-down implementations where we keep the number of free parameters as small as possible. In particular,  \cite{cliff_2021_bbe}  talks about arbitrary-length vectors of competitor-preferences and race-track performance-factors but in our work reported here we keep the length of such vectors to one wherever possible.

Currently BBE models only markets for {\em win bets} (i.e., betting that a particular competitor will win the race, or not): real betting exchange also offer additional markets for any particular event, such as {\em place bets} 
%(e.g., betting that a particular competitor will come in the top $N$, where $N$ often depends on the size of the field, i.e.\ the number of competitors running) 
and {\em each way} bets; 
%(i.e., a combined pair of a win bet and a place bet, on the same race); 
the addition of such supplementary markets in BBE, enabling study of the interplay between win and other markets, is a topic for future work. 

A substantial body of research has been published in various fields that reports on empirical studies of the behavior of actual human bettors (see, e.g., \cite{kanto_etal_1992,swidler_shaw_1995,bradley_2002,jullien_salonie_2008,choi_hui_2014,feess_etal_2015,brown_yang_2016,suhonen_etal_2018,merz_etal_2020}), where there is a common concept of the {\em representative bettor}, i.e.\ an idealisation of the betting behavior of the average bettor, and in which prospect theory (see e.g.\ \cite{tversky_kahneman_1992}) has been a major influence. In this area of the literature, a fair amount of effort has been expended on exploring and explaining the {\em favourite-longshot}\/ bias, where bettors tend to undervalue favourites (outcomes that have short odds, high probability) and overvalue outsiders (outcomes that have long odds, or low probability), a bias that is frequent in betting of all sorts.

For many years, researchers interested in the microstructural dynamics of financial markets (e.g.\ \cite{dejong_rindi_2009}) have analysed high-resolution financial-exchange data and have identified a number of statistical characteristics that are commonly referred to as the ``stylized facts'' of financial market data (e.g.\ \cite{terasvirta_zhao_2010}) such as high kurtosis or slowly decaying autocorrelation. Any synthetic data generator for financial markets would (and should) be judged at least partially on the extent to which it can generate data series that exhibit the same stylised facts.

Unfortunately, as far as we have been able to determine, there is no body of research that identifies similar stylized facts in the time-series data from in-play betting on track-racing events. The closest we have found is the recent PhD thesis by \cite{restocchi_2018_phd}, who analysed data from {\em prediction markets}\/ for political events. As is explained by \cite{cliff_2021_bbe}, prediction markets are close relatives of in-play betting markets, but the two are typically not identical because the opportunity sets available to market participants differ. \cite{restocchi_2018_phd} did find some statistical characteristics in the political prediction market data that bear a reasonable comparison to the stylized facts of financial markets, but the extent to which in-play betting markets exhibit stylized facts, and the nature of those facts if they do exist, is currently unknown: this is a point we return to in Section~\ref{sec:furtherwork}.

In recent years there has been a growing body of research publications exploring the use of statistical approaches, machine learning, and/or artificial intelligence, in betting markets. Various authors have reported mathematical or algorithmic approaches to profitable betting or trading on betting exchanges, often involving machine learning; see, e.g.: \cite{ioulianou_etal_2011,aruajosantos_2014,bunker_susnjak_2019,hubacek_etal_2019,axen_cortis_2020,goncalves_etal_2020, hubacek_sir_2020,wheatcroft_2020}; and \cite{wilkens_2020}. However all of these studies work from databases of historical odds/price time-series from one or more betting exchanges, and none of them report on methods for in-play betting: in this sense, they are comparable to automated methods for identifying buy and sell signals from analysis of historic time-series of daily price-movements in financial markets. Such an approach is perfectly valid, of course, but it gives little or no insight on how best to trade second-by-second in a fast-moving situation such as an in-play betting market for an event that is underway. Furthermore, such an approach fails to capture the closely-coupled feedback loop where events occurring mid-race cause some bettors to alter their opinions, placing new in-play bets, which are then visible to other bettors in the market, causing them bettors to also adjust their positions.

In comparison, the number of research papers reporting on in-play betting is small: as is discussed in more detail in \cite{cliff_2021_bbe}, the publications of \cite{easton_uylangco_2009,tsimpras_2015} and \cite{dzalbs_kalganova_2018} each offer the possibility of exploring in-play betting but either choose not to, or are reporting on systems that are not public-domain open-source SDGs, and hence unlike BBE.

Fundamentally, even the highest-resolution time-series of in-play betting prices for a specific event is only half the story: without a similarly accurate record of how the event itself played out (e.g., second-by-second records of the positions of the competitors on the track), there is simply nothing to correlate the betting activity against: that is, there is no ground-truth.  BBE generates ground truth data by (re-)creating exactly the kind of events that bettors like to bet on. Although in principle BBE could later be extended to incorporate simulations of sporting events such as soccer games or tennis matches, track-race events are a natural place to start because they are relatively straightforward to characterize mathematically, as described in the next section.

\section{The Race Simulator}
\label{sec:racesim}

As currently configured, BBE is an abstract minimal model of some number of bettors interacting via a betting exchange to back and lay bets on the outcomes of racing events. The model is sufficiently minimal and abstract that in principle it could be interpreted as a model of gambling on horse-races or greyhound-racing, two sports in which betting is deeply embedded; or it could equally be interpreted as a model of gambling on races between motor vehicles; or any other type of event where some number of participants are started at the same time and then compete to cross a finish-line first. There is nothing in our model that specifically limits us to one specific type of race, so we will often talk here just of {\em races} and {\em competitors}. 

For any one race, denoted by subscript $r$, some number $n_r$ of competitors compete by racing along a one-dimensional track of specified length $L_r$: the position along the track of competitor $c$ at time $t$ is a real-valued distance denoted by $d_c(t)$, and the state of the race at time $t$ can be summarised by the vector $\vec{d}(t)$ in which the $c^\text{th}$ element is $d_c(t)$ and hence $|\vec{d}|=n_r$. Individual competitors are merely represented as points along the track: they have no physical extent in our model, although they can impede or block one another's progress, as described further below. 

A competitive race starts at time $t=0$, and the clock then continues to run until the last-placed competitor $c$ achieves a position $d_c(t)>= L_r$ -- i.e., the race ends when the slowest competitor crosses the finish line; in-play betting may be specified to end at that at time, or possibly when an earlier condition is met, such as the third-placed competitor passing the line. 

Each competitor's progress within a race is governed by a discrete-time process such that $d_c(t+\delta_t)= d_c(t)+S_c(t)$ where $S_c(t)$ is a function that generates a step-forward for competitor $c$ at time $t$: $S_c(t)>0$ at all times, to ensure that the race will eventually end, and should usually be a stochastic function so that the outcome of the race cannot be determined precisely at the start. For example, using ${\mathbb U}(lo,hi)$ to represent a uniformly distributed random variable over the range $[lo, hi]$, competitor $c1$ might have $S_{c1}(t)={\mathbb U}(10,20)$ while competitor $c2$ might instead have $S_{c2}(t)={\mathbb U}(1,25)$: given the specifications of these two $S$ functions, we can say that one competitor is more or less likely to cross the finish line first on the average, but we cannot say for sure who will win a specific individual race. 

Complete details of the design of the $S_c$ function were given in \cite{cliff_2021_bbe}, to which the reader is referred for the full rationale. Briefly, for a specific race denoted by the subscript $r$, we use ${\cal C}_r$ to denote the set of competitors in that race; and $\vec{f}_r$ to denote the vector of {\em factors}\/ or {\em features}\/ for race $r$, such as whether the track is dry or wet, whether it is flat or undulating, and so on. Each competitor $c$ has a {\em preference}\/ vector $\vec{p}_c$ which indicates its preferred values of the various factors in $\vec{f}_r$, and has a function ${\cal P}_c(\vec{f}_r,\vec{p}_c) \to [0,1] \in {\mathbb R}$ which gives a multiplicative coefficient that can reduce the stochastic step-size taken by that competitor on each timestep by an amount that depends on the degree of mismatch between $\vec{f}_r$ and $\vec{p}_c$. The stochastic step-generating function itself is denoted generally as $\delta(\vec{v}_c)$ where $\vec{v}_c$ is $c$'s vector of parameters for whatever distribution is used for the step-generator, e.g.\ in the example given above, we'd say $\vec{v}_{c1}=[10, 20]^T$ and $\delta() = {\mathbb U}()$. Each competitor also has a {\em responsiveness} function ${\cal R}_c(t,\vec{d}) \to {\mathbb R}^+$ which gives a mechanism for modelling situations in which one competitor might start at a fast pace but slow toward the end of the race, while another might start slow and speed up at the end. The full $S_c$ function is shown in Equation~\ref{eq:S}. 
\begin{equation}
S_c(t,\vec{f}_r,\vec{d}) = 
\begin{cases}
	{\cal R}_c(t,\vec{d}).{\cal P}_c(\vec{f}_r,\vec{p}_c).{\delta}(\vec{v}_c) & \text{if } \Delta_c(t)>\theta_c \\
	{\cal R}_c(t,\vec{d}).\delta_{\text{min}}   & \text{otherwise.}
\end{cases}
\label{eq:S}
\end{equation}  
Where $\theta_c$ is $c$'s threshold distance for being delayed by a slower-running competitor in front of it (i.e., if the distance to the nearest competitor in front is more than $\theta_c$ then $c$ is not delayed by that competitor); and $\Delta_c(t)=d_{i^+}(t)-d_c(t)$ is the distance to the nearest competitor ${i^+}\neq c$ who is in front of $c$, i.e.: 
\[
i^+ = \underset{i \in {\cal C}_r}{\arg\min}(\Delta_c(t), \forall i : d_i(t)>d_c(t)),
\] 
and $\delta_{\text{min}} = \min (S_c(t-\delta_t,\vec{f}_r,\vec{d}),S_i(t-\delta_t,\vec{f}_r,\vec{d}))$, which means that if $i^+$ is too close in front of $c$, then $c$'s step-size becomes limited by $i^+$'s step size, only if $i^+$ is running slower than $c$.

In our simulation, each individual bettor $b_i$ makes predictions about the outcome of a race and bets on the basis of those predictions. Intuitively, the accuracy of an individual bettor's predictions can be situated on a continuum from making equiprobable random choices over the space of possible  outcomes for a particular race (thereby totally ignoring all available information about the nature of the race and about each of the competitors) through to a god-like omnisciently rational bettor who has perfect information on all factors that contribute to the outcome of the race. One way of distributing the population of bettors along this continuum is to initially make each bettor form equiprobable estimates of the likelihood of each outcome for a race, and then to randomly allocate each bettor some number $d$ of ``dry-run'' trials: in any one dry-run, the race is simulated and that bettor uses the outcome of that simulation to revise its estimate of what the outcome will be when the race actually takes place. A bettor with $d=0$ remains a purely random bettor; a bettor with $d=1$ has one trial's worth of data to go on, which is better than nothing but is not as good as $d=10$ or $d=100$; in the limit, as $d \to \infty$, the trial-outcome information that is available to an individual bettor is so extensive that accurate estimates of the probability of each possible outcome for the race can readily be made using elementary frequentist statistics. This is not to say that we think real bettors do this, rather it is a straightforward way of endowing the bettors in our model with an element of rationality. But, as we discuss in Section~\ref{sec:bettors}, BBE also needs {\em irrational} bettors.

\section{The Betting Exchange}
\label{sec:exchange}

Algorithmically, the betting exchange's matching-engine itself is largely a simple matter of ensuring accurate book-keeping. For each competitor in a particular race, a record is kept of all the back-bets on that competitor, and all the competitor's lay-bets, that have been received by the exchange: the internal record of each bet includes the arrival-time of the bet at the exchange, the identity of the bettor, and the amount wagered. Once a bet is received and recorded at the exchange, it can be cancelled by the bettor if it is not yet matched with a counterparty, but matched bets cannot be cancelled. Bettors can submit more than one bet into the market for a particular event. The arrival-time matters because multiple bets at the same odds are processed in time-priority order. The matching process pairs up buys and lays: when a lay at a particular price arrives, it is matched with the oldest unmatched back at that price; and when a back arrives it is matched with the oldest unmatched lay at the same price. A bet for a large stake can be fully matched against multiple bets of the opposite direction for smaller stakes, with the large-stake bet treated internally as being split into multiple smaller separate bets on the exchange. Partial matches remain active, held at the exchange awaiting the subsequent arrival of a matching counterparty bet. All bets unmatched at the close of betting expire and the stakes are returned to the bettors.

The ``market'' for any one competitor $c$ in a specific race is formed by aggregating across all back bets of the same odds, and across all lay bets of the same odds, to calculate the total amount of money wagered as backs or lays at each specific odds, and that is then displayed as the set of odds and total stakes for $c$, in $c$'s row of the overall ``market'' table for that particular race: see \cite{cliff_2021_bbe} for further illustration of this. 
When the event ends, stakes are collected from the accounts of bettors who lost their bets, and the funds are then distributed to those bettors who placed winning bets. Because the exchange is a platform, taking neither side of the bet, it makes its money by charging a small percentage commission fee (e.g.\ 5\%) on winnings; losing bets are not charged.

The BBE matching-engine is the most straightforward component: the functionality of a real-world betting exchange is well documented; and so there is relatively little latitude or room for creativity in the implementation of this component of the BBE simulator, which is why it does not take up much of the discussion here.  Indeed, it could be plausibly argued that the BBE exchange module is not a simulation of a betting exchange; it {\em is} a betting exchange (that is, the core matching engine in BBE is not an {\em abstraction} of the real thing, but is instead an {\em instance} of the real thing). 

There is an awful lot more latitude when it comes to modelling the other major component in BBE, the bettors, as discussed in the next section. 

\section{Modelling Bettors}
\label{sec:bettors}

As far as we have been able to determine, there is very little research literature on the behaviour of human bettors wagering on in-play markets for track-racing events, and we have found no academic papers at all that describe agent-based models of in-play markets on betting-exchanges.\footnote{It is famously difficult to prove a negative but, for the record, we searched Google Scholar, arXiv.org, and SSRN.com using all the relevant keywords that we could think of, and our searches gave no useful results.} The model bettors we describe here are novel, in the sense that we know of no comparable work in the research literature, but they are also preliminary -- a sey of exploratory first steps. All of the model bettors described below were first introduced in \cite{cliff_2021_bbe}, which gives further discussion of their rationale, and further details of their design. 

Fundamentally, any set of strategies for an artificial betting agent can be arranged along a spectrum, a partial ordering, from wholly irrational to wholly rational. A wholly irrational bettor would make a wild guess, while a wholly rational bettor would bet according to the best information available to it. Any real betting exchange is likely to have participants with varying degrees of rationality, so it is important in BBE to have bettor-agents that vary from wild guessers (analogous to the ``noise traders'' used in models of financial markets) to those that try to make the most well-informed and educated estimate of who will win. The latter class of bettor-agent, the most rational ones in BBE, have already been described above in Section~\ref{sec:racesim}: their estimate of the outcome of the race comes from aggregating the results from some number $d$ of ``dry-runs'' of the race simulator -- we refer to these as {\em Rational Predictors}\/ with $d$ dry-runs as RP($d$) bettors, where the higher the value of $d$ the more accurate the bettor's prediction of the outcome is expected to be. The other classes of bettor introduced by \cite{cliff_2021_bbe}, with varying degrees of rationality, are as follows: 

\begin{itemize}

\item LinEx (Linear Extrapolator): The way in which track-racing has been abstracted in BSE, i.e.\ the modelling of the race as each competitor's progress along a one-dimensional number-line racetrack, means that in any one race each competitor's progress along the track can be treated as a time-series of distance measurements, and the introductory end of the vast literature on time-series analysis can then be mined for creating some minimal but plausible in-play betting strategies. One example is referred to within BBE as the ``LinEx'' strategy, because it involves linear extrapolation: a LinEx bettor estimates the current speed of each competitor at each timestep by taking the arithmetic mean of that competitor's stepsize over the past $N$ seconds, and assumes that these speeds will each remain constant for the rest of the race; then, working from each competitor's current position on the track and the estimate of their current speed, LinEx calculates a prediction of which competitor will cross the line first.

\item
LW (Leader Wins): this bettor's view of the outcome of the race is that whichever competitor is currently in the lead will go on to win.

\item
UD (Underdog): this bettor predicts that the second-placed (P2) competitor will win, so long as the distance to the race-leader is less than some threshold distance $D$: if the P2 racer falls behind by more than that, the Underdog bettor switches prediction to the P1 competitor. 

\item
BTF (Back The Favourite): this class of in-play bettor monitors the distribution of stakes in the market and predicts that the winning competitor will be the one that currently has the lowest odds, i.e.\ the market's favourite.

\item
RB (Representative Bettor): a bettor agent that is programmed to behave in ways consistent with research results on human betting behaviors: although there is comparatively little literature covering human behavior in in-play betting, there is \cite{brown_yang_2016} who note that, in the hurry and heat of the moment, humans tend to choose stakes that are nonuniformly distributed across the space of possible amounts but instead cluster on multiples of 2, 5, or 10; and the well-known {\em favorite-longshot} bias is also coded into BBE's RB Bettors. The literature concerning representative bettors is surveyed and discussed further in \cite{cliff_2021_bbe}.

\item
ZI (Zero Intelligence): these bettors choose a competitor at random and assume it will win -- these are the ``noise traders'' of a betting market. 

\end{itemize}

Populating BBE with a suitably large number of these bettors, with randomly-varied values for their parameters (such as $D$ in {\em Underdog}, or $N$ in {\em LinEx}) gives sufficiently rich variation in opinion while the race is underway that the dynamics of the in-play market is plausibly nontrivial, as is illustrated by the sample results shown in the next section.

\section{Three Implementations of BBE}
\label{sec:implement} 

The focus in our work thus far has been to establish three independent implementations of the BBE model, where each implementation takes a ``minimum viable product'' (MVP: see e.g.\ \cite{ries_2011}) approach, prioritising the establishment of a complete end-to-end flow of data, such that each MVP implementation has all the key components in place and has the ability to generate data of the type and scale that BBE was originally intended to deliver. Once the MVP is up and running (the situation we are in at the time of writing this paper), and the source-code is released into the public domain, we are then in a position for us or others to refine and extend the model, improving the match between the data generated by BBE and the data that is generated by real betting exchanges. 

There are two main motivations for producing multiple independent implementations as we have done here: the tradeoff between accessibility and specialization; and the desire to independently replicate results. 

The significance of the accessibility-specialization tradeoff comes from the observation that when implementing any simulation model there is often a tradeoff between technical {\em accessibility} (i.e., how easy it is for a non-expert programmer to understand what is going on) and technical {\em specialization} (i.e., the extent to which advanced techniques are employed to enhance performance). A specialised implementation might run very fast, but would probably require advanced programming skills that are not widely distributed in the community of researchers for whom we expect BBE to be of interest, and so would most likely prove hard for other researchers to adapt and extend. Nevertheless, specialized implementations offer the appeal of potentially using advanced computing approaches such as GPGPU (General-Purpose computing on Graphical-Processing Units -- the specialised silicon chips built for displaying computer graphics, with hundreds of simple computers or ``cores'' all working in parallel, which can give massive speed-ups when correctly programmed), or asynchronous multi-threaded code execution (which models the parallel and asynchronous nature of real exchanges more faithfully than a traditional single-threaded execution mode).    

The motivation of independent replicability comes from the simple desire to be surer of our results than if we only had a single implementation: given the multiple moving parts and the compounded nonlinearities in the BBE model, there are many opportunities for things to go wrong, for honest mistakes to be made in the implementation. If we only had a one implementation, then identifying any problems in that implementation would be made difficult by the lack of any reference point, no other implemenattion to compare it to. If we had two implementations and their results agree, that is appealing but if they disagree then it is not obvious which of the two is in error and which is not; however with three implementations there is a reasonable expectation that two versions might be in agreement while the third is not, indicating which needs fixing. 

Our three implementations are referred to as {\em STP}\/ (Single-Threaded Python, as documented in full by \cite{lausoto_2021_MEng}); {\em MTP}\/ (Multi-Threaded Python, as documented in full by \cite{keen_2021_MEng}), and {\em MCM}\/ (Multi-Core Mixed-language, as documented in full by \cite{hawkins_2021_MEng}). As is signalled by their names, both the STP and MTP implementations are written in {\em Python}, and the MCM implementation is a mix of Python and the specialist GPGPU programming language {\em OpenCL}. 

\begin{figure}[t]
\begin{center}
\includegraphics[width=0.95\linewidth]{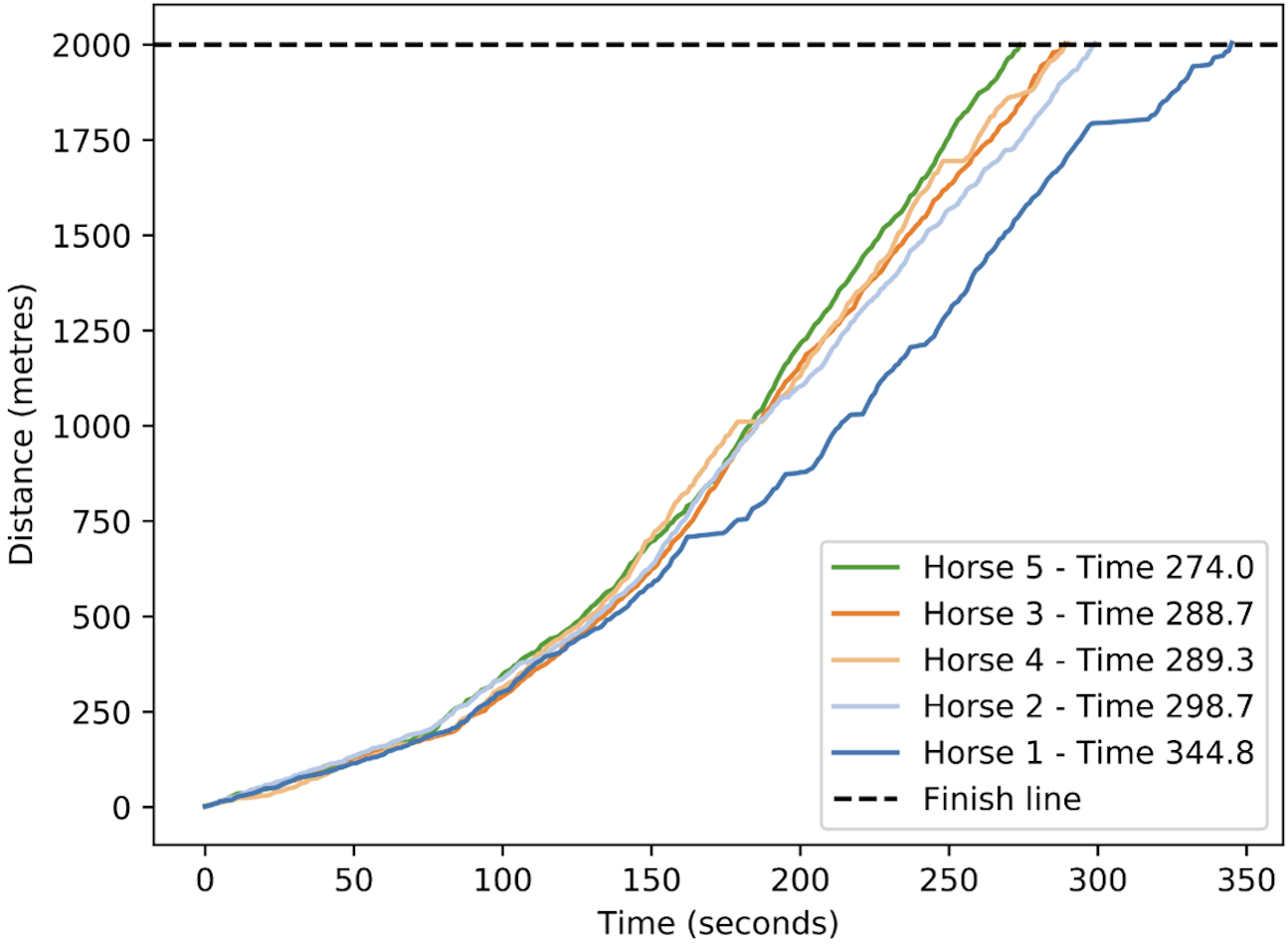}
\end{center}
\caption{Space-time plot for a five-horse race in BBE. The horizontal axis is time in seconds, and the vertical axis is distance travelled in metres: this race is over 2000m and lasts less than six minutes. The competitor called {\em Horse1}\/ starts the race as favourite but falls behind and eventually finishes last, while {\em Horse5}\/ starts with the longest odds but eventually wins. The parameter values for this race were chosen to create an illustrative race with some dramatic turns of fortune occurring: this gives the swings in opinion among the bettor population illustrated in Figure~\ref{fig:RACE1-bettors}; this is not intended as a model of any real race.}        
\label{fig:race1plot}
\end{figure}

Figure~\ref{fig:race1plot} shows illustrative BBE race data generated by the STP implementation of \cite{lausoto_2021_MEng}, for a single five-competitor race over 2000m, lasting less than six minutes: the progress of the race is visualised as a plot of distance over time for each competitor. The competitor named {\em Horse1}\/ starts off as the favourite to win, but ends up coming last. 
 
While Figure~\ref{fig:race1plot}  shows the results from a single $n_r=5$ simulated race, if we run that same simulation repeatedly, $R$ times, with a different random-seed on each run, then that will give us an estimate of the probability mass function (PMF) over the discrete space of possible finish-order outcomes: there are $n_r!$ possible finish-order outcomes. Such PMF estimates from the three simulators can be compared using an appropriate nonparametric significance test such as Kruskal-Wallis to indicate whether the differences between their outcomes are explainable as noise or as consistent variation: in this way the outputs of the race-simulator components of the three implementations can be rigorously compared and aligned with each other. 
 
\begin{figure}
\begin{center}
\includegraphics[width=0.95\linewidth]{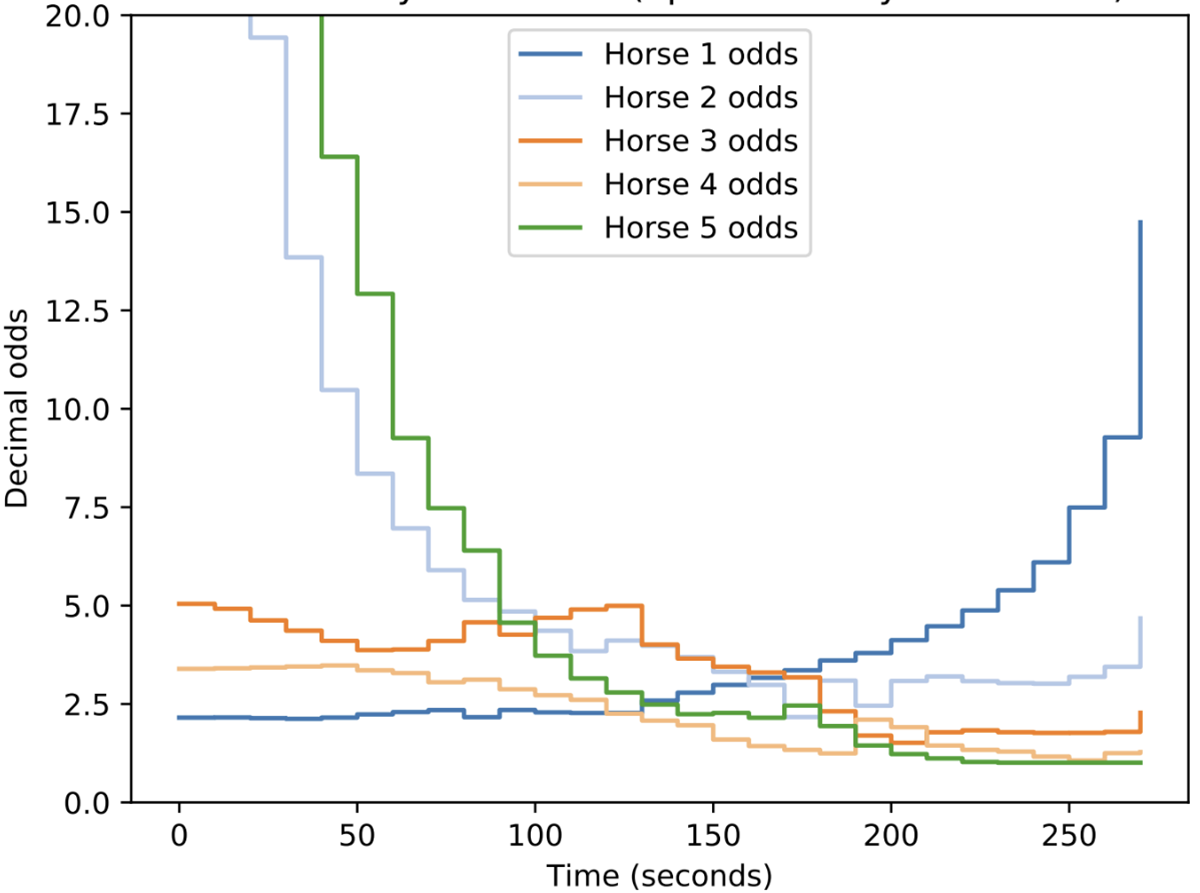}
\end{center}
\caption{Changes in the sentiment of a single RP(d) bettor over the duration of the race illustrated in Figure~\ref{fig:race1plot}. The horizontal axis is time, and the vertical axis shows the decimal odds assigned by that bettor to each horse: the bettor re-evaluates the odds on all horses every 10 seconds, giving the time series a manifestly stepped appearance. As can be seen, Horse 1 starts out as the favourite (i.e., has lowest odds) but as it falls behind the bettor assigns it ever higher odds; similarly, {\em Horse5}\/ starts as the  outsider, but its odds shorten as the race progresses.}
\label{fig:RACE1-bettors}
\end{figure}

Figure~\ref{fig:RACE1-bettors} shows a summary of the market opinion of one RP(d) bettor as the race in Figure~\ref{fig:race1plot} unfolds, and clearly shows how the bettor's opinion (expressed as decimal odds of winning) shifts as the race unfolds. Again, although the stochasticity in the system will mean that any single run from one BBE implementation will likely differ from a run with the same initial conditions executed by one of the other BBE implementations, repeating a large number of trials with different random-number-generator seeds for each trial will generate enough data to argue persuasively that the overall distribution of outcomes within the simulated population of bettors is consistent over the three implementations (or not).  

Figures~\ref{fig:race1plot} and~\ref{fig:RACE1-bettors} come from our STP implementation of BBE, as described in \cite{lausoto_2021_MEng}: this is the least specialised, and most technically accessible, of our three implementations: it is a conveniently conventional single-threaded Python script, and hence should present the fewest barriers to understanding and alteration/extension by non-expert programmers. Our other two implementations each require considerably more technical skill to work with, but the extra technical specialisation in our MTP and MCM implementations brings other advantages. 

Our Multi-Threaded Python (MTP) implementation is described fully in \cite{keen_2021_MEng}, and uses the native multithreading capability of Python3, which means that each bettor-agent in the model is assigned its own {\em thread} (i.e., its own independent virtual processor) and so the set of processes modelling the bettors in BBE all run in parallel and asynchronously. This is important because recent results from models of automated trading systems in financial markets have shown that whether the simulation of traders in the market is single-threaded or multi-threaded can make a major difference on the outcome of the simulations, with the single-threaded approximation to parallelism giving results that are contradicted by the much closer approximation to parallelism that is provided by a multi-threaded implementation (see \cite{rollins_cliff_2020,cliff_rollins_2020}). When it comes to modelling and evaluating the profitability of automated in-play betting strategies, it is reasonable to expect that similar concerns will be relevant: any real-world deployment of an automated in-play betting strategy would have to operate in a fully asynchronous and parallel fashion, and our MTP implementation is a manifestly better approximation of such a real-world deployment than our STP version is, but it comes with the cost that editing and debugging multi-threaded programs requires considerably more specialised coding skills than are needed for working with the STP version.  

Many of the use-cases that BBE is intended for require very many race simulations to be conducted. For example, in a five-minute race with $B$ rational bettors using the RB(d) approach explained in this paper, each with $d=50$, and each updating their private estimates of the outcome of the race once per second, the implementation of BBE will need to run one simulation of the actual race itself, and a further $15000\times B$ ``dry-run'' simulations for the RB(d) bettors; and that is just for one race. Using BBE as the platform for an automated search/optimization process (as described in more depth in the next section) might very plausibly require thousands of different simulated race-events to be run, each with its own particular set of characteristics and conditions (i.e., each with its own $\vec{f}$ factor-vector), and so the number of individual race simulations that need to be executed can quickly rise to be in the millions. Thankfully, many of these simulations are independent from most of the others, and so they are what is known technically as {\em embarassingly parallelizable}: they can be executed in parallel on $P$ different processors, for a $1/P$ speed-up. To exploit this embarassingly parallelizable nature of BBE usage, \cite{hawkins_2021_MEng} has developed the MCM (multi-core multi-language) implementation of BBE which splits out the race simulations onto the cores of a GPU: the race simulation code is written in OpenCL, with the rest of BBE written in Python. This is a highly specialized implementation, and hence is correspondingly low in accessibility, but it gives huge performance speed-ups, of roughly three orders of magnitude: a bulk simulation of a large batch involving many races on the MCM implementation might take only a few seconds, whereas running the same batch of simulations on the STP or MTP simulators might take thousands of seconds.

\begin{figure}
\begin{center}
\includegraphics[width=0.95\linewidth]{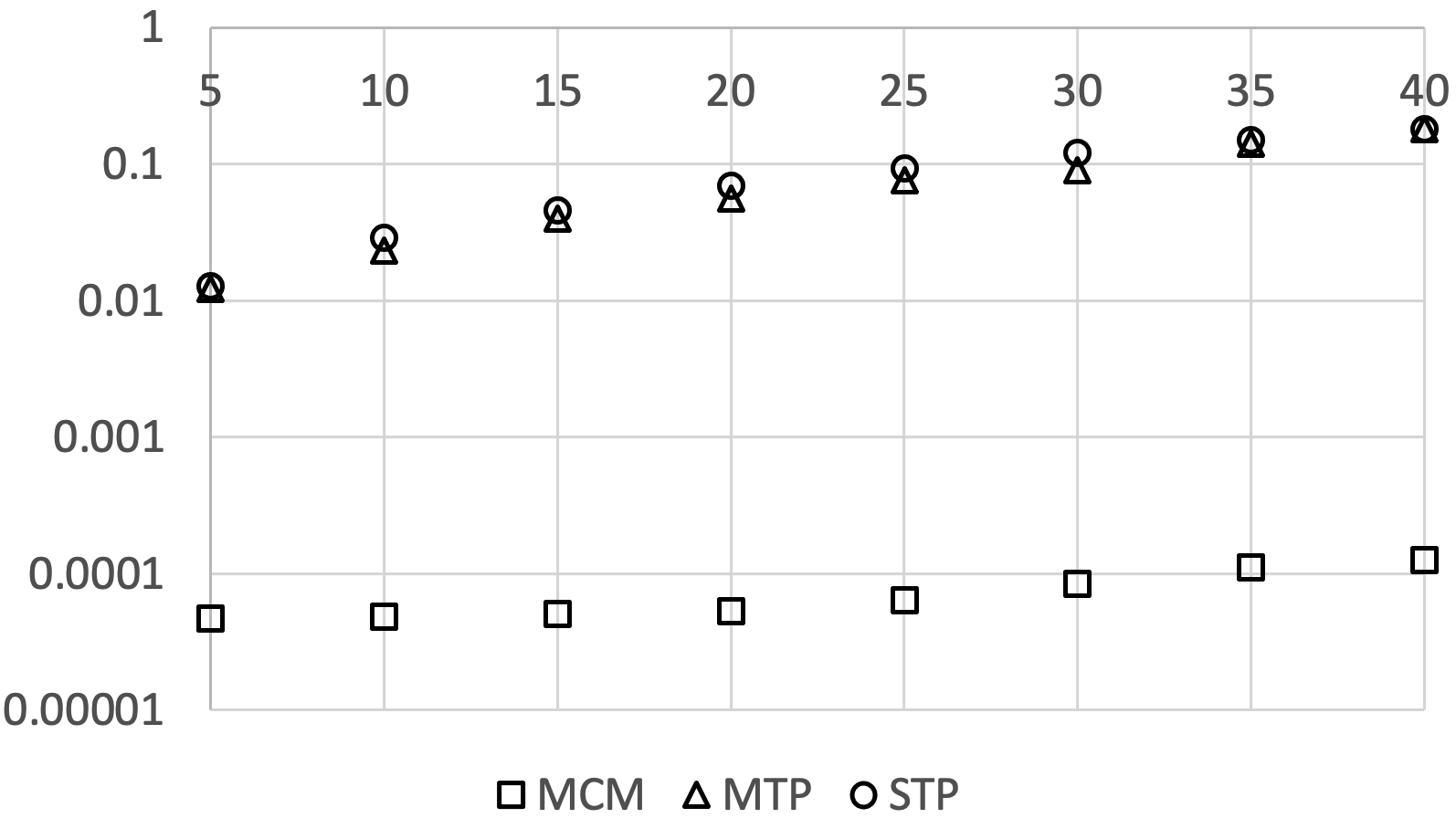}
\end{center}
\caption{Comparison of run-time performance scaling of our three implementations, as the number of competitors in the race is varied while all other parameters are held constant. The horizontal axis is the number of competitors in a single race, on a linear scale; the vertical axis is log-scaled and shows the mean wall-clock time elapsed (in seconds) to simulate a single race. Points labelled STP come from our Single-Threaded Python implementation; points labelled MTP come from our Multi-Threaded Python implementation; and points labelled MCM come from our Multi-Core Multi-language model, where the races are simulated in OpenCL on a GPU. To generate these data, the parameters for the three implementations were each set to give a race distance of 2km with competitors running at speeds that give an expected finish time of six minutes or less, and with the race timestep $\delta_t$=1s. STP data points were generated by measuring the total elapsed execution time for running 1,000 i.i.d.\ race simulations and calculating the mean; similarly, the MCM data points are mean values from runs of 100,000 i.i.d.\ simulated races on a 640-core {\sc Nvidia} GeForce GTX 1050 GPU. Overall, the STP and MTP results are so close as to be essentially identical, while the MCM model runs more than $1000\times$ faster than the other two models.}
\label{fig:compare_horses}
\end{figure}

To illustrate the performance differences between our three implementations, Figure~\ref{fig:compare_horses} shows scaling data of ``wall-clock'' time elapsed for our three implementations as the number of competitors is increased and as all other parameters are held constant, and identical across the three implementations. As can be seen, the STP and MTP results are so close to each other as to be almost identical: this is because the multi-threading in the MTP implementation parallelises the simulated bettors, not the races. And, most obviously, there is a difference of approximately three orders of magnitude between the execution speed of the STP implementation (which is the most accessible and easy to edit for non-experts) and the MCM implementation (which is the least accessible, given the technical specialization that has been deployed to achieve this huge speed-up): that is, MCM runs in roughly one thousandth of the time taken by STP and MTP. 

\begin{figure}
\begin{center}
\includegraphics[width=0.8\linewidth]{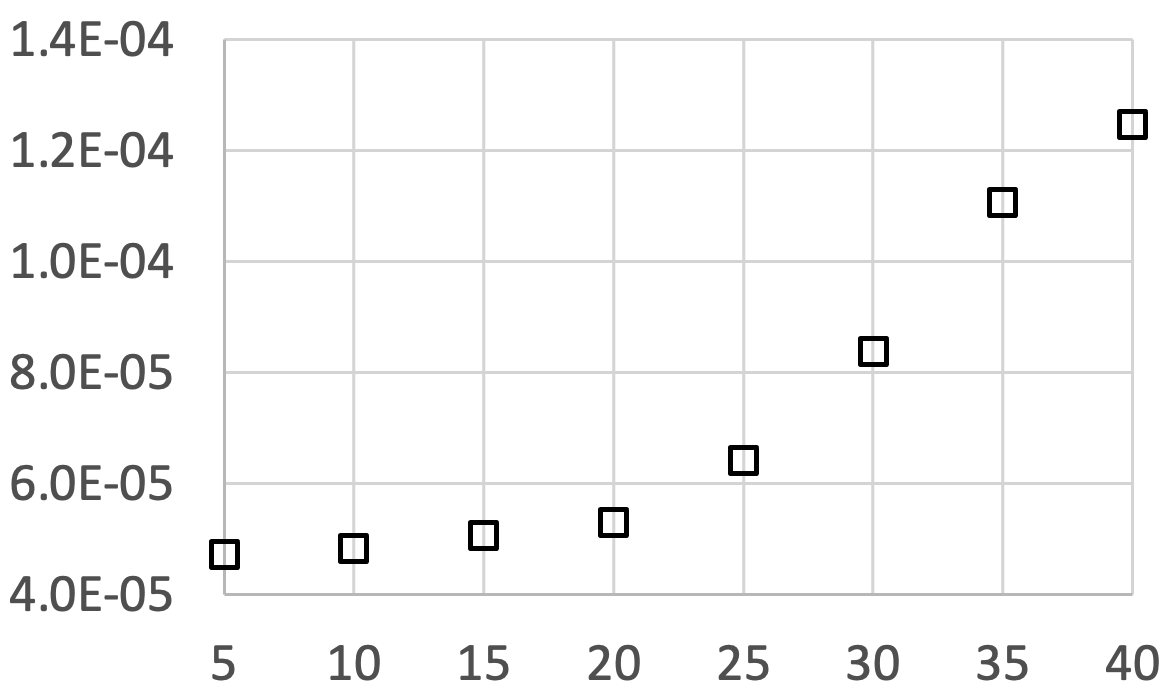}
\end{center}
\caption{Close-up view of the MCM data-points plotted in Figure~\ref{fig:compare_horses}: the axes here are the same as in Figure~\ref{fig:compare_horses} but the vertical axis is now linearly scaled. As can be seen, the increase in runtime is nonlinear, because of constraints dependent on the width of the SIMD path on the GPU that executes the MCM OpenCL. The coefficient of variation in these data is very small, typically around 0.005: at this scale, error-bars plotted at $\pm 1$ standard deviation would be invisible.}
\label{fig:mcm_closeup}
\end{figure}

Figure~\ref{fig:mcm_closeup} shows a close-up view of the MCM data from Figure~\ref{fig:compare_horses}: as is clear from this graph, the growth in runtime on MCM is nonlinear overall, as the number of instructions being executed in parallel will be limited by the number of SIMD (single instruction multiple data) lanes available on the GPU chip: once the GPU runs out of SIMD lanes, it will switch from increasing the parallel bandwidth to increasing the number of sequential cycles being executed. However, from a practical perspective this is minor detail, given the huge increase in speed that the MCM approach offers. 

\section{Further Work}
\label{sec:furtherwork}
 
The initial objective in our work was to establish three independent MVP implementations of the BBE model, prioritising the establishment of end-to-end flows of causation and data, such that our MVP-style implementations have all the key components in place and can generate output data of the type and scale that BBE was originally intended to deliver: as we have demonstrated in this paper, that objective has now been met. This then enables the use of BBE as a platform on which automated search and optimisation processes can be based, which is one direction for our future work. For instance, one research direction we are now exploring is the use of evolutionary computing techniques such as genetic algorithms (GAs) and evolution strategies (ESs) to explore very large spaces of possible designs of betting strategies, in the hope that the GA/ES discovers profitable betting strategies. As a proof-of-concept (PoC), we are commencing this work with deliberately small spaces of possible designs that include pre-existing betting strategies that are known to be profitable: by deliberately setting up the PoC such that a known success is possible, we can check that the evolutionary process is capable of ``discovering'' the pre-existing strategies. Once we know that the evolutionary search process does operate successfully, we can greatly expand the search-space, making it unbounded, and observe to see whether new betting strategies can be evolved.  

Another direction for future work that we are currently embarking upon was already mentioned above: now we have the end-to-end data-flow in place, we can work on fine-tuning the details of the BBE implementations so that the statistical characteristics, the ``stylised facts'' of the data produced from BBE are usefully close to the stylised facts of real in-play markets on actual betting exchanges. But to do that we will need first to engage in analysis of raw in-play data from real exchanges because, as far as we have been able to determine, no other researchers have yet published any detailed statistical analyses of the market microstructure of in-play track-race betting markets. 

We intend to report on both of these two strands of further work in future publications. 

\section{Conclusions}
\label{sec:conclusion}

Sections 1 to~\ref{sec:bettors} of this paper provided a heavily compressed summary of  \cite{cliff_2021_bbe}, which surveyed relevant literature and argued for the research opportunity presented by  an agent-based simulation model of a group of bettors interacting via a betting exchange to make ``in play'' back and lay bets on the outcome of a race event, while that event is underway: that model is referred to as BBE. The novel contribution of this paper is to show the results from, and discuss the comparison between, the three independent replications fully documented in \cite{hawkins_2021_MEng,keen_2021_MEng}; and \cite{lausoto_2021_MEng}: we refer the reader to all three of those documents for complete descriptions of the work reported here.  To the best of our knowledge, BBE is the first agent-based simulation model of its kind, in that no other in-race betting-exchange simulators are available as open-source research resources in the public domain. By reporting here on the development and open-source release of three independently produced implementations that we have made freely available to the research community, our software can be used, explored, and extended by researchers with varying levels of appetite for technical specialization. 

We have demonstrated here that BBE can be used to generate fine-grained data-sets on sub-second temporal resolution from arbitrarily large number of simulated races: this enables the very low-cost generation of extremely large synthetic data-sets that can be used for training data-intensive machine learning systems in the search for profitable automated wagering and trading on betting exchanges. Now that our simulators are established at an initial operational level of maturity, future papers will report on the results from generating and using such data to explore active research questions. With the BBE source-code being made available on GitHub, our hope is that other researchers will now use the BBE model as a common platform, facilitating ready replication and extension of results, and hopefully will also contribute to further developing the BBE codebase as required.

%%%%%%%%%%%%%%
%% References
%\bibliographystyle{apa}

%% Specify your .bib file name here, without the extension
\bibliography{../dc_bibliography}

%% End of the document
\end{document}